# Plenoptic microscope based on Laser Optical Feedback Imaging (LOFI)

W. Glastre, O. Hugon, O. Jacquin, H. Guillet de Chatellus and E. Lacot
Laboratoire Interdisciplinaire de Physique
Université de Grenoble, Centre National de la Recherche Scientifique
Grenoble, France
Eric.Lacot@ujf-grenoble.fr

*Abstract*— We present an overview of the performances of a plenoptic microscope which combines the high sensitivity of a laser optical feedback imaging set-up, the high resolution of optical synthetic aperture and a shot noise limited signal to noise ratio by using acoustic photon tagging. By using an adapted phase filtering, this microscope allows phase drift correction and numerical aberration compensation (defocusing, coma, astigmatism …). This new kind of microscope seems to be well adapted to make deep imaging through scattering and heterogeneous media.

## I. Introduction

Making fast 3D images with a good in-depth resolution through both heterogeneous and turbid media have always been a major issue. The problem is double with scattering media: firstly the scattering medium generally attenuates strongly the signal, which decreases the signal to noise ratio (SNR) and secondly the wavefront is highly perturbed, which degrades the point spread function (PSF) of the imaging system and therefore the resolution. Several ways to overcome these problems have been proposed; two main methods aiming at keeping a good optical resolution are actively developed. The first one uses pre-compensation of the wavefront before propagation, to improve the resolution [1]. The second one only uses ballistic photons to make images [2,3]. Our laser optical feedback imaging (LOFI) setup [3], based on optical self-mixing inside the laser cavity, belong to this second family. LOFI has the advantage of providing a self-aligned and very sensitive optical system limited by shot noise [4,5].

In this paper, after a brief reminder of what LOFI is, we present our plenoptic microscope which combines LOFI (based on the ballistic photons detection), synthetic aperture (SA) [6,7] and acoustic photon tagging (APT) [7,8]. This new kind of microscope demonstrates the possibility of making deep imaging through scattering and heterogeneous media at the shot noise level [9].

## II. Synthetic Aperture LOFI

### A. Experimental setup

Fig. 1 shows a description of the LOFI experimental setup. The laser is a microchip laser emitting a power of tens mW at $\lambda = 1064$ nm. This laser has a relaxation frequency $F_R$ in the megahertz range. On its first pass, the laser beam is frequency shifted by a frequency $F_e/2$ where $F_e$ is close to the relaxation frequency of the laser ($F_R \approx F_e$), and then sent to the bi-dimensional target by means of two rotating mirrors. The beam diffracted and/or scattered by the target is then reinjected inside the laser cavity after a second pass in the galvanometric scanner and the frequency shifter. The total frequency shift undergone by the photons reinjected in the laser cavity is therefore $F_e$ which results in triggering relaxation oscillations of the microlaser and in amplifying the sensitivity of the device to the reinjected photons. A small fraction of the output beam of the microchip laser is sent to a photodiode. The delivered voltage is analyzed by a lock-in amplifier at the demodulation frequency $F_e$, which gives the LOFI signal (i.e. the amplitude and the phase of the electric field of the reinjected light). Experimentally, the LOFI images (amplitude and phase) are obtained pixel by pixel (i.e. point by point, line after line) by a full 2D galvanometric scanning. In the case of SA LOFI, the target is located at a distance L from the focal plane of the objective. The image is therefore obtained with a defocused beam. The raw complex image $h_R(x,y)$ must be numerically filtered to realize post focusing with the advantages to obtain images beyond the working distance of the lens [6,7]. In order to increase the image contrast an acoustic transducer can be used to tag (i.e. to shift the laser beam frequency by an additional amount $F_a$) the specific photons which have reached the immersed target

### B. SA Refocusing

When a punctual target is scanned with the defocused beam, one obtains the blurred PSF [6,7]:

$$h_R(L,x,y) \propto G(F_E)\left( \exp\left(-\frac{x^2+y^2}{(\lambda L/\pi r)^2}\right) \exp\left(j2\pi\frac{x^2+y^2}{2L\lambda}\right) \right)^2, \quad (1)$$

where $G(F_E)$ is the LOFI gain [3,5] at the frequency $F_e$ and where r is the laser beam waist in the image focal plane of the objective.

By taking the 2D Fourier transform of this expression, one obtains:

$$H_R(L,\upsilon,\mu) \propto \exp\left(-\frac{\upsilon^2+\mu^2}{(\sqrt{2}/\pi r)^2}\right)\exp\left(-j\frac{\pi L\lambda(\upsilon^2+\mu^2)}{2}\right), \quad (2)$$

where $\upsilon$ and $\mu$ are the spatial frequency coordinates along x and y directions. The right-term of (2) shows the defocus which corresponds to the quadratic phase dependence. To numerically refocus raw images, this phase has to be cancelled by multiplying the signal in Fourier space by the phase filter:

$$H_{filt}(L_{filt},\upsilon,\mu) = \exp\left(j\frac{\pi L_{filt}\lambda(\upsilon^2+\mu^2)}{2}\right) \quad (3)$$

This filter corresponds to the free space retro-propagation transfer function over a distance $L_{filt}/2$ (the factor 2 is due to the round trip configuration of LOFI). After an adapted filtering ($L_{filt}=L$) and inverse Fourier transform in the spatial domain, one has the following final synthetically refocused signal:

$$|h_{SA}(x,y)| = |TF^{-1}(H_R(L,\upsilon,\mu)H_{filt}(L,\upsilon,\mu))| = \exp\left(-\frac{x^2+y^2}{(r/\sqrt{2})^2}\right) \quad (4)$$

In this expression $TF^{-1}$ is the inverse Fourier transform. One can observe that after numerical refocusing, the resolution is equal to $r/\sqrt{2}$ (and not r, due to the round trip configuration of LOFI ), whatever the initial defocus L, is [6,7]. This property can also be used to obtain in a single 2D scan a pseudo 3D image (image of non-plane surface) [9].

Figure 2 shows, an example of a numerical refocusing for a plane target composed of small silica beads of 30-40 µm diameter located behind a π shape hole with both a width and a height of the order of 1 mm.

*C. Acoustic Photon Tagging*

As shown on Fig. 3, there is a drawback to SA concerning the photometric balance. By comparison with the classical LOFI imaging (focalized case), the SA signal power is decreased by a ratio $\pi r^2/S_{SA}$ (ratio between the surface of the laser beam in the image plane of the objective and in the plane of the target) [7]. So we have a signal loss proportional to $L^2$ and we can quickly be limited by optical parasitic reflections (see background on Fig. 3a and 3b) which are above the shot noise limit (the LOFI ultimate limit) [3,5].

In order to eliminate parasitic signal from unwanted reflections, we can use APT which eliminates by filtering the contribution of parasitic reflections and offers real shot noise sensitivity giving access to a greater depth (L). One needs to tag photons just before the target [7, 8]. This tagging is obtained with an acoustic transducer which focuses an acoustic wave in the image plane of the objective (see Fig. 1). This acoustic wave produces a sinusoidal modulation of the pressure (amplitude ~MPa) at the frequency $F_A/2 = 2.25$ MHz.

In order to consider only tagged photons for the detection, the signal can be demodulated (with the lock-in amplifier) at the frequency $F_e - F_A$ instead of previous $F_e$. With the complete setup of Fig. 1, we made an image of a target composed of small silica beads of 30-40 µm diameter located behind a circular hole of 1 mm diameter. This target was placed in a rectangular glass tank filled with diluted milk acting as the scattering medium. Without APT we have $F_e = 4.4$ MHz and with APT, we have $F_A = 4.5$ MHz and $F_e = 8.9$ MHz. As a result, in both cases (with or without APT), the total round-trip frequency shift is equal to 4.4 MHz. By comparing Figs. 3a and 3b with Figs. 3c and 3d we clearly see an improvement with the reduction of the background and therefore an increase of the image contrast (i.e of the image SNR). In both cases the optical resolution is recovered after numerical refocussing.

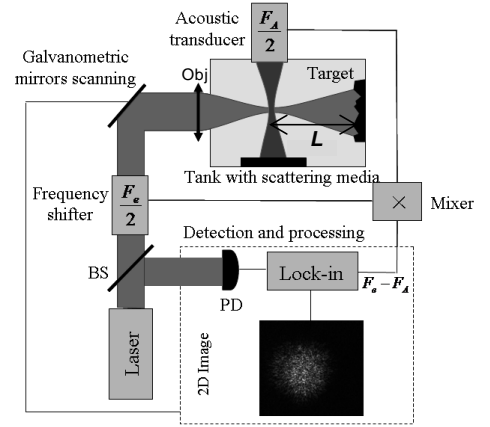

Fig. 1. Schematic diagram of Synthetic Aperture LOFI setup with acoustic tagging. PD, photodiode; BS, beam splitter; $F_e$ and $F_A$ round-trip frequency shifts respectively induced by acousto-optic modulators and acoustic transducer; × frequency mixer. The target is located at a distance L from the focal plane of the objective (Obj).

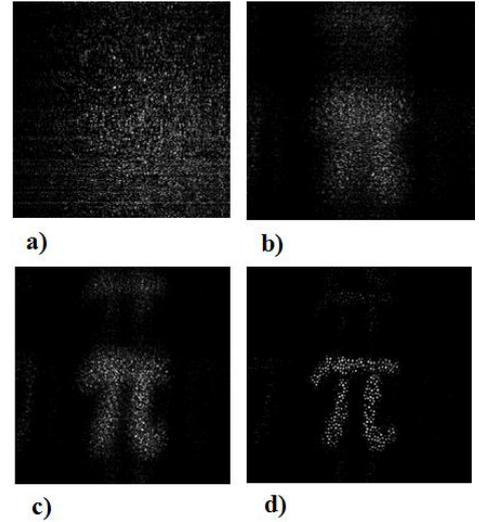

Fig. 2. Example of numerical refocusing. a) Raw image, b) Refocussing image with a filter mismatch ($\Delta L = L_{filt} - L = 4$ cm), c) Filter mismatch ($\Delta L = 2$ cm), d) Adapted filter ($\Delta L = 0$ cm).

*D. Phase drift compensation*

During the LOFI image acquisition, made point by point, line after line, a slow phase drift can occurs, which degrades the final synthetic image (i.e. the refocused image). This drift is mainly due the variations of optical path between the laser and the target due to slight variations of the refractive index of the air (because of temperature or pressure changes). In typical experiment conditions (i.e. without any particular isolation), the phase drift is about 1 radian/minute in our lab, and principally affect the slow direction of the scan. Because our total acquisition time is about 1 minute, we see that this phase noise (which acts like an optical aberration) is very important (of the order of $\pi$) and needs to be corrected before SA processing.

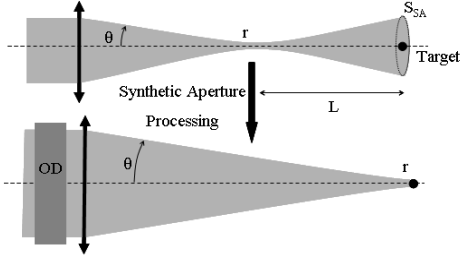

Fig. 3. Effect of Synthetic Aperture operation on the raw acquisition equivalent setup. L is the raw defocus, r the beam waist and θ the numerical aperture. OD = $\log(S_{SA}/\pi r^2)$ is the equivalent Optical Density and $S_{SA}$ the surface of the laser beam after a propagation over a distance L.

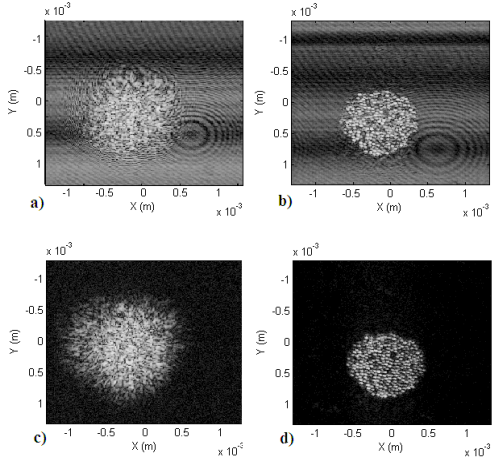

Fig. 4. Comparison of LOFI images, with and without APT. a) and b), images without APT; c) and d), images with APT; a) and c), raw images; b) and d) images after SA processing.

To do so, we propose a simple and efficient solution which consists in taking two raw images. The first one with a quick scan along the horizontal direction and therefore with a phase drift principally along the vertical direction and a second image by inverting the slow and fast acquisition directions. By combining these two images (i.e. by subtracting the two phase images), we can numerically determine the phase drift in any direction (horizontal or vertical) and we can finally recover a corrected "raw" acquisition [6]. By applying SA filtering, one can finally obtain a quasi-perfect synthetic image.

Fig. 5a shows the amplitude of the raw image, when the vertical direction (i.e. Y) is the slow direction of acquisition. Fig. 5b, shows the amplitude of the corresponding synthetic aperture image. This last image clearly shows, that, the slow phase drift causes vertical aberrations and that the synthetic image is blurred in the Y direction. By using the second acquisition (not shown here), we can determined the phase drift (see Fig. 5c) by eliminating vertical phase difference between the two images . Finally, when SA operation is applied to the corrected raw image, we get Fig. 5d, showing that our blur (i.e. aberration) correction method is efficient.

### III. PLENOPTIC LOFI

Aberration compensation is also an important concern for imaging through heterogeneous biological media [10]. To correct aberrations, one possible solution consists in the introduction of adaptive optics resulting in an aberration-free laser spot in the target plane. Spatial light modulators or deformable mirrors can be used in this way [11,12]. Another way to handle that problem is to use a plenoptic detector (i.e. a detector recording simultaneously the position and the direction of propagation of the photons), where the compensation is made by a numerical post-processing of the acquired raw images.

The LOFI setup, which allows to record the feedback electric field (i.e. the amplitude and the wave front of the complex electric field), is therefore a plenoptic imaging setup [9]. To demonstrate aberration compensation in a SA LOFI setup, we can first consider the simplified situation of a constant aberration in the image field (*i.e.* we assume that the raw PSF does not depend on the field. Under this assumption, one obtains in the Fourier space the following transfer function:

$$H_R^{'}(L,\upsilon,\mu) = H_R(L,\upsilon,\mu)H_{aber}(\upsilon,\mu) \qquad (5)$$

In this expression $H_{aber}(\nu,\mu)$ is the plane waves dephasing term responsible for the aberrations. In order to both refocus and correct aberrations, the filter function is turned into:

$$H_{filt}^{'}(L,\upsilon,\mu) = H_{filt}(L,\upsilon,\mu)H_{aber}^{-1}(\upsilon,\mu) \qquad (6)$$

To illustrate the capabilities of the LOFI technique for aberration correction, spherical silica beads of 30-40 μm diameter are imaged by the LOFI setup depicted on Fig. 1. The advantage of this object is that a single bead acts as a punctual reflector (Dirac) located at its centre. The raw PSF is then directly accessible. Contrary to the paraxial lens usually used, an objective introducing large aberrations is now placed in the setup. This objective is a simple plano-convex spherical lens which is voluntarily tilted to induce important asymmetrical aberrations relatively to the optical axis (astigmatism and coma). From raw image, the defocus is first removed which leads to Fig. 6a where only remain higher order aberrations. Then astigmatism and coma are removed leading to Fig. 6b. This final image can be compared to Fig. 6d where the image has been obtained with an aberration-free objective.

To finish, on Fig. 6c, $|h_{SA}(L,x,y)*h_{aber}(L,x,y)|$ is represented with $h_{aber}$ the inverse Fourier transform of $H_{aber}$. This illustrates the astigmatism and coma that are compensated and thus which were initially present in the objective. By comparing Figs. 6a and 6b, an important improvement in the image quality can be observed, which confirms the interest of our numerical aberration compensation technique. However by comparing Figs. 6b and 6d, it is also possible to see that these corrections are not totally perfect and needs to be improved.

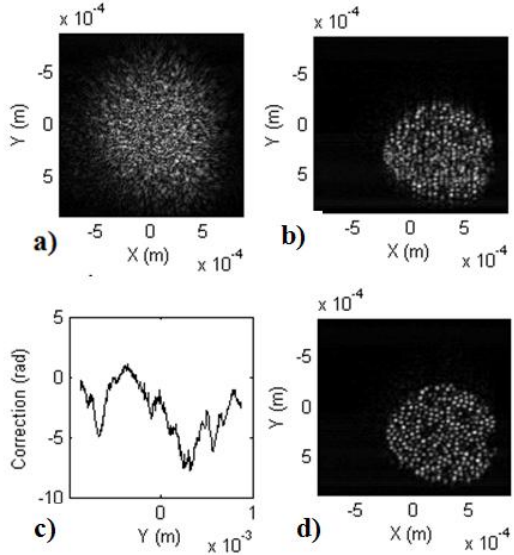

Fig. 5. Correction of the effects of the phase drifts occuring during the raw acquisition. a) Raw image; b) Synthetic image without phase drift correction; c) Phase drift experimentally determined; d) Synthetic image with phase drift correction. One can see that the blur in the Y direction has disappeared.

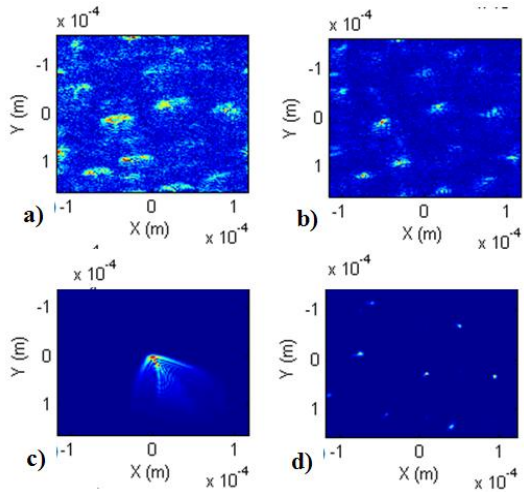

Fig. 6. Example of partial abberation compensation of the plenoptic LOFI microscope. Image a) is obtained after numerical refocusing and b) with both refocusing and numerical aberration compensation. c) is the PSF used for filtering which shows the aberrations which are compensated. d) Perfect refocused image obtained with an abberation-free objective (this last figure obtained with an aberration-free objective has not be acquired on the same zone of the field: this explains why no correlations on the placement of the beads can be observed).

## IV. CONCLUSIONS AND PERSPECTIVES

In this paper, an overview of the properties of a plenoptic microscope based on LOFI technique is made. This plenoptic microscope combines the high resolution of optical synthetic aperture and the high sensitivity of a self-mixing heterodyne technique called LOFI. This association enables to overcome the limitation of the accessible depth due to the objective working distance by numerical refocusing. This plenoptic microscope has also been coupled with APT to reach shot noise sensitivity. Other interesting properties of the plenoptic microscopes, such as numerical aberration compensation by using an adapted filtering and numerical removing of an unwanted phase drift have also been demonstrated. All these properties are possible at the price of a slow point by point galvanometric scanning and of the degradation of photometric performances when the target is not in the objective's imaging focal plane. In the future, we plan to extend this work by implementing a correction of local aberrations beyond local defocus. Such an improvement will enable aberrations-free images through both scattering and heterogeneous media, which paves the way to biological sample imaging.